\documentclass[amssymb, amsmath, aps, prd, showpacs, nofootinbib]{revtex4}
\usepackage{bm}
\usepackage{graphicx}
\expandafter\ifx\csname package@font\endcsname\relax\else
 \expandafter\expandafter
 \expandafter\usepackage
 \expandafter\expandafter
 \expandafter{\csname package@font\endcsname}
 \fi
\usepackage{hyperref}

\newcommand{\bb}{\textbf}

\newcommand{\be}{\begin{eqnarray}}

\newcommand{\bit}{\begin{itemize}}
\def\bkt#1{\left(#1\right)}
\def\bkts#1{\left[#1\right]}

\newcommand{\ee}{\end{eqnarray}}
\def\eg{\textit{e.g.} }
\newcommand{\eit}{\end{itemize}}

\def\etal{\textit{et al.} }

\newcommand{\ff}{\phantom{t}}

\def\gtsima{$\; \buildrel > \over \sim \;$}

\def\iee{\textit{i.e. }}
\newcommand{\ii}{\textit}

\def\lab{\label}

\def\lt{\left}
\def\ltsima{$\; \buildrel < \over \sim \;$}

\newcommand{\nn}{\nonumber}

\def\pr{\prime}
\def\Qc{Q^{(c)}}

\def\re#1{(\ref{#1})}
\def\rt{\right}

\def\simlt{\lower.5ex\hbox{\ltsima}}
\def\simgt{\lower.5ex\hbox{\gtsima}}
\def\sub#1{_{\mbox{\scriptsize{#1}}}}

\begin{document}

\title{Cosmological perturbations in models of coupled dark energy}
\author{Sirichai Chongchitnan}
\email{siri@astro.ox.ac.uk}
\affiliation{Oxford Astrophysics, Denys Wilkinson Building, Keble Road, Oxford, OX1 3RH, United Kingdom.}

\begin{abstract}
Models in which dark energy interacts with dark matter have been proposed in the literature to help explain why dark energy should only come to dominate in recent times. In this paper, we present a dynamical framework to calculate cosmological perturbations for a general quintessence potential and interaction term. Our formalism is built upon the powerful phase-space approach often used to analyse the dynamical attractors in the background. We obtain a set of coupled differential equations purely in terms of dimensionless, bounded variables and apply these equations to calculate perturbations in a number of scenarios. Interestingly, in the presence of dark-sector interactions, we find that dark energy perturbations do not redshift away at late times, but can cluster even on small scales. We also clarify the initial conditions for the perturbations in the dark sector, showing that adiabaticity is no longer conserved in the presence of dark-sector interactions, even on large scales. Some issues of instability in the perturbations are also discussed.

\end{abstract}

\date{October 2008}
\pacs{98.80.Cq}

\maketitle

\section{Introduction}

Recent datasets from a range of cosmological experiments have convincingly suggested that roughly 25\% of the Universe is in the form of weakly interacting cold dark matter, whilst over 70\% of the Universe is filled dark energy with negative pressure that accelerates  the cosmic expansion \cite{wmap5, riess, kowalski, percival}. Despite these observational breakthroughs, the theoretical origin of this \ii{dark sector} remains poorly understood. Although most datasets can be explained by the $\Lambda$CDM model, in which dark energy is simply the cosmological constant, the data cannot yet rule out the possibility that dark energy may be dynamical, and that there may be undiscovered interactions in the dark sector (see \cite{copelandreview} for a review).

The simplest model of dark-sector interactions involves dark matter and dynamical dark energy in the form of quintessence, a slowly-rolling scalar field evolving along a potential $V(\phi)$ \cite{ratrapeebles,caldwell}. It is well-known that certain classes of $V(\phi)$ possess a desirable dynamical property  such that field configurations from a wide range of initial conditions evolve towards a common attractor. Furthermore, if there exists dark-sector interactions that allow dark matter to continuously decay into dark energy, it may be possible to evoke such interactions to help explain the so-called ``cosmic coincidence" problem, \iee why dark energy only comes to dominate in recent times \cite{amendola,farrar, holden}. 

The presence of dark-sector interactions would clearly affect the evolution of cosmological perturbations, which in turn affects the details of large-scale structure formation. Since specific cosmological details are clearly model-dependent, it would be useful to have a general prescription with which cosmological perturbations could be calculated given any model of quintessence and the dark-sector coupling. This is the primary aim of this paper.

There have been many investigations into the perturbations in coupled dark energy scenarios (see, for examples, \cite{hn, jussi, amendola2, brookfield, lee, olivares, bean2,mainini, koivisto}). However, our approach is different   as the calculation of perturbations will be entirely combined with the powerful phase-space formalism often used to analyse the dynamical attractors in the background \cite{halliwell,wands, copeland,bohmer}. As a result, the perturbation equations derived will be expressed in terms of dimensionless variables which are bounded, hence allowing a large range of initial conditions to be explored. We shall demonstrate the use of this technique with a few examples, which reveal interesting behaviour of the perturbations in the presence of dark-sector interactions. 

Some conventional views on cosmological perturbations are that matter density perturbations evolve adiabatically on large scales, and that dark-energy perturbations simply redshift away at late times. In the simplest models of quintessence, dark energy may cluster, but this is significant only on scales so large  that it is irrelevant for astrophysical processes. We shall see that, in the presence of dark-sector interactions, all these statements are no longer true. Moreover, in some cases, instabilities can develop in the perturbations even though the background evolves smoothly.

Throughout this work, we set $c=1$ but keep the gravitational constant $G$ explicit.

\section{Background evolution}
We assume a flat, homogeneous Friedmann-Robertson-Walker universe with metric
\be ds^2 = a^2(-d\tau^2+dx_i dx^{i}),\ee where $\tau$ is the conformal time related to the cosmic time $t$ by $dt=a d\tau$. We also assume that dark energy is a quintessence $\phi$ with potential $V(\phi)$, which is non-negative and at least twice differentiable in $\phi$. 

The Friedmann equation is
\be H^2 &=& {\kappa^2\over3}\bkt{\rho_r+\rho_c+\rho_b+V+{1\over2}\dot\phi^2},\ee
where $\kappa=\sqrt{8\pi G}=m\sub{Pl}^{-1}$ and $\rho_r, \rho_c, \rho_b$ are the energy densities of radiation, cold dark matter and baryons (an overdot indicates a derivative with respect to  $t$ unless stated otherwise). The energy density and pressure of quintessence are given by
\be \rho_\phi = {1\over2}\dot\phi^2+V,\qquad p_\phi = {1\over2}\dot\phi^2-V.\lab{energy}\ee 

The energy conservation equation
\be \dot{\rho_i}+3H\rho_i(1+w_i)=0,\ee is assumed to hold for radiation ($w_r=1/3$) and baryons ($w_b=0$). The energy exchange in the dark sector can be represented by
\be \dot{\rho_c}+3H\rho_c&=&Q,\lab{exchange1}\\ \dot{\rho_\phi}+3H\rho_\phi(1+w_\phi)&=&-Q,\lab{exchange}\ee where the coupling $Q$ is a function of $\phi$. In particular, if $Q<0$, then dark matter instantaneously decays into dark energy. The reverse occurs when $Q>0$. Nevertheless, it is important to note that the sign of $Q$ does not by itself determine the \ii{overall} effect of dark-matter--dark-energy conversion. For instance, it is possible for a model with $Q<0$ to end up with {less} dark-energy density at late times than a model with $Q>0$, as we shall see later.

Using Equations \re{energy}, Equation \re{exchange} becomes a modified Klein-Gordon equation
\be \ddot{\phi}+3H\dot{\phi}+V^\pr(\phi)&=&-\Gamma,\lab{kg}\ee where we define 
\be Q\equiv\Gamma\dot{\phi}.\ee 

The evolution of the various densities can be easily studied by introducing the dimensionless variables \cite{halliwell,wands, copeland}
\be x\equiv {\kappa\dot\phi\over \sqrt{6}H}, \quad y \equiv {\kappa\over H}\sqrt{V\over 3},\quad z \equiv {\kappa\over H}\sqrt{\rho_r\over 3}, \quad u \equiv {\kappa\over H}\sqrt{\rho_c\over 3},\quad v \equiv {\kappa\over H}\sqrt{\rho_b\over 3},\lab{dimless1}\ee
and, in addition, a dimensionless interaction variable
\be \quad \gamma \equiv {\kappa\Gamma\over\sqrt{6}H^2}.\lab{dimless2}\ee 
The Friedmann equation then becomes the constraint
\be x^2+y^2+z^2+u^2+v^2=1,\ee which means that the phase-space is simply the surface of the sphere ${\Bbb S}^4$.

Next, differentiating each variable with respect to $N=\ln a$ gives
\be {d\ln H\over dN}&=&-{3\over 2}\bkt{1+x^2-y^2+{1\over3}{z^2}},\lab{ode1}\\
{dx\over dN} &=&-\gamma+\sqrt{3\over2}\lambda y^2-x\bkt{3+{d\ln H\over dN}},\lab{ode2}\\
{dy\over dN} &=&-\sqrt{3\over2}\lambda xy-y\bkt{d\ln H\over dN},\lab{ode3}\\
{dz\over dN} &=&-z\bkt{2+{d\ln H\over dN}},\\
{du\over dN} &=&{\gamma x\over u}-u\bkt{{3\over2}+{d\ln H\over dN}},\\
{dv\over dN} &=&-v\bkt{{3\over2}+{d\ln H\over dN}}\lab{ode6},\ee
where the $\lambda$ is the `roll' parameter defined by
\be\lambda \equiv-{V^\pr(\phi)\over\kappa V(\phi)}.\ee Note that in this definition, $N=0$ at the present time and decreases into the past. If $\lambda$ is constant, then $V$ is just the exponential potential $V_0 e^{-\kappa\lambda\phi}$. In general, any positive potential can be parametrized by a dynamical $\lambda(N)$. 

In summary, given a potential $V(\phi)$, a coupling function $Q(\phi)$ and initial conditions for $(H,x,y,z,u,v)$, the set of coupled differential equations \re{ode1}-\re{ode6} completely specifies the evolution of the background cosmology.

\section{Perturbations in a general model}
 
While the dynamics of the background with dark-sector interactions have been well-investigated using phase-space techniques,  the perturbations have so far been calculated by solving the perturbed field equations separately \cite{hn, hn2, jussi, bertschinger,brookfield, bean}. This renders the two approaches somewhat disparate. 

In fact, it is possible to view quintessence models as trajectories in the phase-space $(x,y,u,v,w)$, while simultaneously studying the perturbations in the various energy densities along the trajectories. The main advantage of such an approach is that information about the attractors in the phase-space can be easily used to calculate the perturbations at (or near) various attractors. Initial conditions for the various densities are also easier to deal with since they correspond to finite regions in the phase-space. 

To demonstrate our technique, we consider a system with only dark matter and quintessence, and derive equations in the phase-space $(x,y,u)$ governing the evolution of dark-matter density contrast, $\delta_c$, and that of the quintessence perturbation, $\delta\phi$, defined by 
\be \delta_c\equiv{\delta\rho_c\over{\bar\rho}_c}, \quad \phi\equiv\bar\phi+\delta\phi,\ee
where an overbar indicates a background quantity. 

\subsection{Perturbation equations}

As a starting point, we briefly follow the approach of Hwang and Noh \cite{hn} in their rigorous investigation of `gauge-ready' forms of the perturbation equations for interacting fluids.

Without fixing any gauge, the general scalar perturbations of the spatially flat FRW metric are given by 
\be ds^2=-a^2(1+2\alpha)d\tau^2-2a^2\beta_{,i}d\tau dx^{i}+a^2\bkt{(1+2\varphi)\delta_{ij}+2\hat{\gamma}_{,ij}}dx^idx^j,\ee 
where  $\alpha,\beta,\varphi,\hat{\gamma}$ are variables with small amplitudes. The components of the energy-momentum tensor are
\be T^0_0=-\rho,\quad T^0_\mu=-{1\over k}(\rho+p)v_{,\mu},\qquad T^\nu_\mu=p\delta^\nu_\mu,\lab{components}\ee
where the total energy density $\rho$ decomposes into the background value ($\bar{\rho}_c+\bar\rho_\phi$) and the perturbed part ($\delta\rho_c+\delta\rho_\phi$) and $k$ is the wavenumber. The total pressure $p$ and velocity $v$ can be similarly decomposed. Overbars will now be omitted unless necessary.

Hwang and Noh also derived perturbation equations in the `comoving' gauge in which the dark matter peculiar velocity $v_c$ vanishes. The variables $\delta_c$ and $\delta\phi$ were shown to satisfy
\be \ddot{\delta_c}+2H\dot{\delta_c}-{\kappa^2\over2}\rho_c\delta_c&=&\kappa^2\bkt{2\dot{\phi}\delta\dot{\phi} -V^\pr\delta\phi} -{1\over a^2}{d\over dt}\left\{{a^2 \over \rho_c}\lt[\Gamma\dot{\phi}\delta_c-\dot{\phi}\delta\Gamma-\Gamma(\delta\dot\phi+3H\delta\phi)\rt]\right\} \nn \lab{hneq1}\\
&+&\bkt{3\dot{H}+2\kappa^2\dot\phi^2-{k^2\over a^2}}{\Gamma\over\rho_c}\delta\phi.\\
\delta\ddot\phi+3H\delta\dot\phi+\bkt{{k^2\over a^2}+V^{\pr\pr}}\delta\phi&=&\dot{\phi}\dot{\delta_c}+{\Gamma\dot\phi^2\over\rho_c}\delta_c -(1+{\dot\phi^2\over\rho_c})\delta\Gamma -{\Gamma\dot\phi\over\rho_c}\delta\dot\phi-\dot\phi{d\over dt}\bkt{\Gamma\delta\phi\over\rho_c}\nn \\&+&2(\ddot\phi+3H\dot\phi){\Gamma\delta\phi\over\rho_c}\lab{hneq2}.\ee  Unless specified otherwise, the comoving gauge will be used for the rest of this paper. By changing the time variable to $N$ and using the dimensionless variables \re{dimless1}-\re{dimless2} and their evolution \re{ode1}-\re{ode6}, the perturbation equations \re{hneq1}-\re{hneq2} can be transformed into a dimensionless form


\begin{widetext} 
\be\frac{d^{2}\delta_{c}}{dN^{2}}&+&A\frac{d\delta_{c}}{dN}+B\delta_{c}=C\frac{d\delta\gamma}{d N }+D\delta \gamma+ E\frac{d \left(\kappa \delta \phi \right)}{d N }+ F(\kappa\delta\phi), \lab{master1}\\
\frac{d ^{2}\left(\kappa \delta \phi \right)}{d N ^{2}}&+& \hat{A}\frac{d \left(\kappa \delta \phi \right)}{d N }+ \hat{B}(\kappa\delta\phi) =\hat{C}\frac{d \delta_{c}}{dN}+\hat{D}\delta_{c}+\hat{E}\delta\gamma, \lab{master2}\ee
which is a set of coupled second order differential equations, additionally sourced by the perturbation in the dark-sector interaction $\delta\gamma$. The coefficients are given in terms of $(x,y,u)$ and $\gamma$ by

\be A&=&\frac{d{\ln H} }{dN}+2,\qquad\quad\quad\ff   B=-\frac{3}{2} u ^{2}+\frac{1}{u ^{2}}\left[2   x \frac{d \gamma }{d N }+4   x\gamma  +4   \frac{d {\ln H} }{d    N } x\gamma +\sqrt{6} \lambda  y^{2}\gamma -2  \gamma^{2}\right]-{8\gamma^2x^2\over u^4}\nn,\\
C&=&\frac{2 x}{u^{2}},\qquad\qquad\qquad\quad\ff    D= \frac{1}{u ^{2}}\left[4  x +4  \frac{d {\ln H} }{d   N }x -4\gamma +\sqrt{6} \lambda y^{2} \right]-{8\gamma x^2\over u^4},\nn\\
E&=&2 \sqrt{6} x +\frac{\sqrt{6} }{u^{2}}\left[\frac{5}{3} \gamma +{2\over3}\frac{d {\ln H} }{d   N } \gamma +\frac{1}{3}\frac{d\gamma}{dN}-2 \frac{ \gamma^{2} x }{u ^{2}}\right],\nn\\
F&=&3y^{2}\lambda +\frac{\sqrt{6}}{u^{2}}\left[5 \gamma +4\gamma\frac{d{\ln H}}{d  N}+\frac{d\gamma}{dN}+4\gamma x^2-\gamma\eta y^2-{2\gamma\over3}\bkt{k\over aH}^2\right]\lab{coeff}\\
&+& \bkts{2\gamma\lambda y^2-{2\over3}\sqrt{6}\gamma^2-2\sqrt{6}\gamma x-{2\over3}\sqrt{6}\gamma x\frac{d\ln H}{dN}-{\sqrt{6}x\over3}{d\gamma\over dN}}{2\gamma\over u^4}+{4\sqrt{6}\over3}{\gamma^3x^2\over u^6},\nn\\
%
%
\hat{A}&=&  {d\ln H\over dN}+{4\gamma x\over u^2}+3 ,\nn\\ \hat{B}&=& 3\eta y^2+\bkt{k\over aH}^2+\bkts{3\gamma+2\gamma{d\ln H\over dN}+\frac{d\gamma}{dN}-{2\gamma^2x\over u^2}}{2x\over u^2}+{4\gamma^2\over u^2}-2\sqrt{6}{\gamma\lambda y^2\over u^2}\nn,\\
\hat{C}&=&\sqrt{6}x, \qquad\quad\quad\hat{D} = 2\sqrt{6}{\gamma x^2\over u^2},\quad\qquad\quad\hat{E}= -\sqrt{6}(1+{2x^2\over u^2}),\nn\ee
\end{widetext}
where $d\ln H/dN=-3(1+x^2-y^2)/2$. We also define the curvature \be \eta \equiv {V^{\pr\pr}\over\kappa^2V}.\ee These equations completely describe the perturbations along the trajectories in the phase-space $(x,y,u)$. They are completely general and valid for any quintessence potential, parametrized by $\{\lambda,\eta\}$, and any interaction rate $\gamma$. Notice that most terms in the coefficients depend on $\gamma$, its derivatives, or its perturbation $\delta\gamma$.

\subsection{Perturbed interaction}

It remains to show how $\delta\gamma$ may be calculated from a given interaction term $Q$ which appears in the energy exchange equations \re{exchange1}-\re{exchange}.


First, we note that the introduction of $Q$ amounts to a modification of the Einstein equation, and so one must be able to associate $Q$ with some covariant entity, say, $Q_\mu$, which satisfies
\be\nabla_\nu T^{(c)\nu}_{\mu}=Q^{(c)}_{\mu},\lab{momentum}\\
\nabla_\nu T^{(\phi)\nu}_{\mu}=Q^{(\phi)}_{\mu},\\
Q^{(c)}_\mu=-Q^{(\phi)}_\mu,\ee
where the total energy-momentum tensor $T^{\nu}_{\mu}= T^{(c)\nu}_{\mu}+T^{(\phi)\nu}_{\mu}$ satisfies the conservation $\nabla_\nu T^{\nu}_{\mu}=0$. The vector $Q^{(c)}_\mu$ can be decomposed into the background and perturbed parts as \cite{hn}
\be Q^{(c)}_0&=& -a\bkt{\bar{Q}^{(c)}(1+\alpha)+\delta Q^{(c)}}\lab{deltaq},\\
Q^{(c)}_{i}&=& J^{(c)}_{,i}. \lab{j}\ee 
Inserting these into \re{momentum} and using \re{components}, we find 
\be\bar{Q}^{(c)}=Q, \qquad \alpha= {J^{(c)}\over\rho_c}.\lab{j2}\ee
The exact forms of $\delta \Qc$ and $J^{(c)}$ can be identified by perturbing the vector $Q^{(c)}_{\mu}$ and comparing with Equations \re{deltaq}-\re{j2}. We shall demonstrate this procedure for a few  examples in the next section. 

Next, recall that earlier we introduced $\Gamma ={Q/\dot\phi}$. By writing $\Qc_\mu=-\Gamma\phi_{,\mu}$ and making the perturbations $\Gamma\rightarrow\Gamma+\delta\Gamma$ and $\phi\rightarrow\phi+\delta\phi$, Equations \re{momentum}-\re{j2} imply 
\be\delta\Gamma&=&\bkt{\delta\Qc-\Gamma(\delta\dot\phi-\alpha\dot{\phi})}/\dot{\phi},\lab{deltaG}\\
\alpha&=& -{\Gamma \delta\phi\over\rho_c}.\lab{alpha}\ee
Finally, having obtained $\Gamma$ and $\delta\Gamma$, multiplying them by $\kappa/\sqrt{6}H^2$ gives the dimensionless variables $\gamma$ and $\delta\gamma$, which appear in the main perturbation equations \re{master1}-\re{master2}. 

Note also that in the case without interaction $(\Gamma=0)$, we find $\alpha=0$, which is just the synchronous gauge condition, as observed in \cite{hn2}. The synchronous and comoving gauges do not coincide if there is any dark-sector interaction.





\section{Applications}

We now apply Equations \re{master1}-\re{master2} to calculate the perturbations in two models with simple interaction terms found in the literature. By setting $\gamma=0$, we also easily obtain perturbations in the case without interaction. 

In addition to the dark-matter density contrast $\delta_c$ and the field perturbation $\delta\phi$, we can also calculate the dark-energy density contrast $\delta_\phi\equiv\delta\rho_\phi/\rho_\phi$, which quantifies the clustering of quintessence, as follows. The energy-momentum tensor of quintessence is
\be T^{(\phi)}_{\mu\nu}=\phi_\mu\phi_\nu-g_{\mu\nu}\bkt{{1\over2}\phi_\sigma\phi^\sigma+V}.\ee
By perturbing the 00-component and equating with $T^{(\phi) 0}_{\ff\ff\ff0}=-(\rho_{\phi}+\delta\rho_{\phi})$, we find
\be \delta\rho_{\phi}=\dot\phi\delta\dot\phi-{\dot\phi}^2\alpha+V^\pr\delta\phi.\ee
Dividing the above by $\rho_\phi$ and using \re{energy} and \re{alpha}, we arrive at a useful expression for the dark-energy density contrast
\be \delta_\phi={1\over3(x^2+y^2)}\bkts{\sqrt{6}x{d(\kappa\delta\phi)\over dN}+\bkt{2\sqrt{6}\gamma {x^2\over u^2}-3\lambda y^2}\kappa\delta\phi}.\lab{qcluster}\ee

%
%

\subsection{Initial conditions}

If there exists an attractor with a sufficiently large basin of attraction in the phase-space, then the evolution of the various background densities are generally insensitive to the initial conditions for $(x,y,u)$. However, the evolution of the perturbations $\delta_c$ and $\delta\phi$ may not necessary be insensitive to the initial conditions in $\delta_c$ and $\delta\phi$. These conditions determine the relative amplitudes of the various growing and decaying modes, and, therefore, must be carefully specified. 

In this paper, we shall impose the adiabatic initial condition, which specifies that the entropy perturbation between dark matter and quintessence \ii{and} that intrinsic to the quintessence vanish at early times. Let $\mathcal{S}$ and $\mathcal{I}$ denote the relative and intrinsic entropy perturbations respectively. The adiabatic initial condition implies that
\be  \mathcal{S}&\equiv&{\delta\rho_c\over\dot\rho_c}-{\delta\rho_\phi\over\dot\rho_\phi}=0,\\ \mathcal{I} &\equiv&{\delta\rho_\phi\over\dot\rho_\phi}-{\delta p_\phi\over\dot p_\phi}=0.\ee
In terms of the dimensionless variables and field perturbations, the above conditions correspond to
\be \mathcal{S}\equiv0\ff&\Leftrightarrow&\ff \delta_c= {3u^2-2\gamma x\over6xu^2(\gamma+3x)}\bkts{\sqrt{6}x{d(\kappa\delta\phi)\over dN}+\bkt{2\sqrt{6}\gamma {x^2\over u^2}-3\lambda y^2}\kappa\delta\phi},\lab{adia1}\\ \mathcal{I}\equiv0\ff&\Leftrightarrow&\ff -\sqrt{6}(\gamma+3x)(\kappa\delta\phi)= \bkts{\sqrt{6}x{d(\kappa\delta\phi)\over dN}+\bkt{2\sqrt{6}\gamma {x^2\over u^2}-3\lambda y^2}\kappa\delta\phi}.\lab{adia2} \ee Combining these conditions, we see that adiabaticity implies that the initial matter density contrast is proportional to the initial field perturbation, with \be \delta_c=\bkt{2\gamma x-3 u^2 \over\sqrt6 x u^2} \kappa\delta\phi.\lab{adia3}\ee

It is well-known that in the case with no interaction ($\gamma\equiv0$), adiabaticity is conserved on large-scales \cite{malquarti,bartolo}. However, interactions in the dark sector can source isocurvature perturbations even for large-scale modes that are initially adiabatic. For example, using the adiabatic conditions, the expression for the derivative of $\mathcal{S}$ is
\be {d\mathcal{S}\over dt}=\bkts{-{1\over3\sqrt{6}x}\bkt{k\over aH}^2+f(\gamma)}(\kappa\delta\phi)+\bkts{g(\gamma)-{1\over3x}-{2x\over3u^2}}\delta\gamma,\lab{Sprime}\ee
where $f(\gamma)=g(\gamma)=0$ whenever $\gamma=0$. Thus, we see that even in the large-scale limit, $d{\mathcal{S}}/dt\neq0$ unless the interaction and its perturbation vanish. A similar expression for $d{\mathcal I}/dt$ can be found, albeit more complicated. Consequently, one cannot consistently insist that both $\mathcal{S}$ and $d\mathcal{S}/dt$ vanish initially, as was done in \cite{amendola2}. The key point here is that in the presence of dark-sector interactions, adiabaticity is generally no longer conserved on large scales. 

\subsection{Example I: No interaction}\lab{nocouplingscaling}

First, we check that the system \re{master1}-\re{master2} correctly reproduces well-known results in the case without interaction. Setting $\gamma=\delta\gamma\equiv0$ gives

\be\frac{d^{2}\delta_{c}}{dN^{2}}&+&\bkt{\frac{d{\ln H} }{dN}+2}\frac{d\delta_{c}}{dN}-\frac{3}{2} u ^{2}\delta_{c}= 2\sqrt{6}x\frac{d \left(\kappa \delta \phi \right)}{d N }+ 3y^2\lambda(\kappa\delta\phi),\lab{noint1}\\
\frac{d ^{2}\left(\kappa \delta \phi \right)}{d N ^{2}}&+& \bkt{\frac{d{\ln H} }{dN}+3}\frac{d \left(\kappa \delta \phi \right)}{d N }+ \lt[3\eta y^2+\bkt{k\over aH}^2\rt](\kappa\delta\phi) =\sqrt{6}x\frac{d \delta_{c}}{dN}.\lab{noint2}\ee

Phase-space analyses \cite{copeland,ng} show that during the matter-dominated era, there exists a global attractor- the so-called scaling solution, whenever $\lambda>\sqrt{3}$. During the scaling regime, we have 
\be x=y=\sqrt{3\over2\lambda^2}, \quad u^2=1-{3\over\lambda^2}, \quad {d\ln H\over dN}=-{3\over2}.\ee  

To see how the perturbations evolve during the scaling regime, we insert the coordinates of the scaling solution into \re{noint1}-\re{noint2} and seek solutions of the form $\delta_c\propto \delta\phi\propto\exp(pN)(=a^p)$. In the large-scale limit $(k\ll aH)$, the system can be solved analytically. The solutions are
\be p = 1,\quad -{3\over2},\quad -{3\over4}\pm{3\over4}\sqrt{{24\over\lambda^2}-7}.\lab{scalinguncoupled}\ee
These solutions agree with those found in \cite{hn2}. The first two modes are the well-known growing and decaying mode in the absence of quintessence. The quintessence induces two additional modes whose nature depends on $\lambda$. Quintessence density contrast \re{qcluster} also grows as $a^{p}$ at the scaling solution.


Phase-space analyses also reveal another attractor for $\lambda<\sqrt{3}$, at which quintessence completely dominates the cosmic densities. In this regime, 
\be x={\lambda\over\sqrt{6}},\quad y=\sqrt{1-{\lambda^2\over6}}, \quad u=0, \quad {d\ln H\over dN}=-{\lambda^2\over2}.\lab{another}\ee  

By proceeding as above, one finds that the resulting coupled system can also be solved analytically. In the large-scale limit, there is now one scale-invariant mode ($p=0$), and three additional modes corresponding to the solutions of the cubic equation
\be p^3-(\lambda^2-5)p^2+\bkts{{\lambda^4\over4}-{9\lambda^2\over2}+6}p+{\lambda^4\over2}-3\lambda^2=0.\lab{cubic}\ee
In the limit of a flat potential with $\lambda\rightarrow0$ (the so-called `skater' models \cite{sahlen2}) or potentials with very gentle slope, the perturbations consist of i) a constant (scale-invariant) mode $\delta_c\propto\delta\phi$ = constant, ii) a decaying mode $\delta_c\propto a^{-2}$ and iii) a decaying mode $\delta\phi\propto a^{-3}$. This behaviour of $\delta_c$ during the dark-energy dominated epoch agrees with standard results (see \eg \cite{dodelsonbook}). By setting $\lambda=\gamma=0$ in \re{qcluster}, we also find that the quintessence density contrast is exponentially damped towards zero in this limit. Thus, quintessence cannot cluster during a dark-energy dominated era. This, however, is not necessarily true if there are interactions in the dark-sector, as we shall see shortly.


\bigskip

We now turn to the case with nonzero interaction. In each of the examples that follows, we summarise the background dynamics in terms of attractors in the phase-space. We then calculate the perturbations as the trajectories approach the various attractors.

\subsection{Example II: $Q=\beta H\rho_c$}


The interaction of the form
\be Q= \beta H\rho_c.\lab{Q1}\ee
appeared in an earlier work of Lima \etal \cite{lima} as a particle-creation mechanism during inflation, but has subsequently been studied in the context of dark energy by a number of authors (\cite{jussi,billyard,zimdahl}, amongst others). It is not straightforward to see what vector $Q_\mu$ correctly gives rise to this form of $Q$. In a recent work, Valiviita \etal \cite{jussi} carried out an analysis using 
\be Q^{(c)}_\mu=Q u_\mu^{(c)},\lab{qproposed}\ee where $u_\mu^{(c)}$ is the dark matter velocity \be u_\mu^{(c)}=a\bkt{-1-\alpha,(v_c+\beta)_{,i}}\lab{pertu}.\ee
Physically, \re{qproposed} means that the rate of interaction of dark matter is greatest along the direction of its velocity.

By perturbing this vector using \re{pertu} and equating with \re{deltaq}, we deduce that 
\be\delta\Qc=Q \delta_c.\ee 
In terms of the dimensionless variables, we have
\be \gamma &=& {\beta u^2\over 2x},\lab{gamma1}\\
\delta\gamma&=&{\beta u^2\over 2\sqrt{6}x^2}\bkt{\sqrt{6}x\delta_c-{d(\kappa\delta\phi)\over dN}-\beta(\kappa\delta\phi)}.\ee

Substituting all this into \re{master1}-\re{master2} gives the coupled system 

\be\frac{d^{2}\delta_{c}}{dN^{2}}&+&\bkt{\frac{d{\ln H} }{dN}+2}\frac{d\delta_{c}}{dN}-\frac{3}{2} u ^{2}\delta_{c}= \bkts{2\sqrt{6}x+{\beta(3-\beta)\over\sqrt{6}x}}\frac{d \left(\kappa \delta \phi \right)}{d N }+\bigg[3y^2\lambda
+2\sqrt{6}\beta x\nn\\&+&{\beta(9-2\beta)\over \sqrt{6}x}{d\ln H\over dN}+{5\beta(3-\beta)\over\sqrt{6}x}+{\lambda y^2\beta(\beta-3)\over2x^2}-{u^2\beta^2(\beta-3)\over 2\sqrt{6}x^3}-{\beta\over\sqrt{6}x}\bkt{k\over aH}^2\bigg](\kappa\delta\phi),\lab{masterbeta1}\\
\frac{d ^{2}\left(\kappa \delta \phi \right)}{d N ^{2}}&+& \bkts{\frac{d{\ln H} }{dN}+3+\beta\bkt{1-{u^2\over 2x^2}}}\frac{d \left(\kappa \delta \phi \right)}{d N }+ \bigg[3\eta y^2+\bkt{k\over aH}^2\nn\\ &+&\beta^2\bkt{{u^2\over x^2}-1}+\beta\bkt{ \frac{d{\ln H}}{dN}   +3-{3\sqrt{6}\lambda y^2\over 2x}  }\bigg](\kappa\delta\phi) =\sqrt{6}x\frac{d \delta_{c}}{dN}-{\beta\sqrt{6}u^2\over2x}\delta_c.\lab{masterbeta2}\ee
Note that by setting $\beta=0$, we recover the system \re{noint1}-\re{noint2} for the non-interacting case. 

\subsubsection{At the scaling solution}

If $\lambda$ and $\beta$ satisfy the conditions \cite{bohmer}
\be|\beta|\leq3,\qquad -2\beta\leq\bkt{3-\beta\over \lambda}^2\leq3-\beta,\lab{conditions}\ee
then the scaling solution exists, with 
\be x= {3-\beta \over\sqrt{6}\lambda},\quad y=\bkt{{(3-\beta)^2\over 6\lambda^2} +{\beta\over3}}^{1\over2},\quad u= \bkt{1-{(3-\beta)^2\over3\lambda^2}-{\beta\over3}}^{1\over2},\quad {d\ln H\over dN} ={\beta-3\over2}.\lab{coords1}\ee This is a global attractor and hence the growth rate of perturbations are insensitive to initial conditions in $(x,y,u)$. When the phase-space is projected onto the $x$-$y$ plane, the locus of the scaling solution is the hyperbola
\be y^2-\bkt{x-{\lambda\over\sqrt6}}^2=1-{\lambda^2\over6}. \lab{locus}\ee
 
At the scaling solution, the dark-energy density is given by
\be \Omega_\phi=x^2+y^2={(3-\beta)^2+\beta\lambda^2\over 3\lambda^2}.\lab{omphi}\ee
As mentioned earlier, it is not the sign of $Q$ but, in this case, the expression \re{omphi} that determines the overall late-time effect of the coupling on the dark-energy density. Indeed, if $\Omega_\phi|_{\beta=0}$ is the dark-energy density in the case without coupling, then 
\be \lambda^2+\beta-6\begin{cases}>0 & \Rightarrow \Omega_\phi>\Omega_\phi|_{\beta=0},\\ <0 & \Rightarrow \Omega_\phi<\Omega_\phi|_{\beta=0}. \end{cases}\lab{cases}\ee 
The effect of the interaction satisfying the first case in \re{cases} allows more dark energy to be produced at late times compared with the non-interacting case. This may be of some use in alleviating the coincidence problem. 

\begin{figure}
\centering
\includegraphics[width=5.6cm, angle = -90]{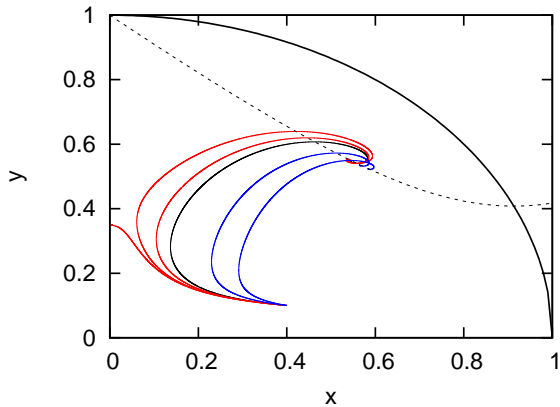}
\caption{Trajectories in the $x-y$ plane for the exponential potential $V=V_0 e^{-\kappa\lambda\phi}$ with $\lambda=\sqrt5$ and interaction term of the form $Q=\beta H\rho_c$. From left to right, the trajectories correspond to $\beta=0.04, 0.035, 0.02, 0, -0.1$ and $-0.2$. The behaviour near the $y$ axis seen in the left-most trajectory is discussed in the text. The dotted hyperbola is the locus of scaling solutions [Equation \re{locus}].}
\label{figbeta}
\end{figure}

The effect of the coupling of the type \re{Q1} on the background is summarised in Figure \ref{figbeta}, which shows the phase space projected onto the $x$-$y$ plane.  In the figure, we set $\lambda=\sqrt{5}$ and $\beta$ ranging from $-0.2$ to 0.04 (right to left). These parameters satisfy the conditions \re{conditions} for the existence of the scaling solution. From a common set of initial conditions, trajectories are evolved towards the scaling solutions, which vary very slightly in position along the locus \re{locus}, shown by the dashed line. Increasing the value of $\beta$ effectively lengthens the path towards the attractor and pushes the trajectories to explore regions with smaller $x$. Physically, this means that the interaction with positive values of $\beta$ effectively acts as an opposing force to the motion of the field. Similarly, the interaction with $\beta<0$ accelerates the field towards the scaling regime.

To solve for the perturbations during scaling, one might try substituting the coordinates \re{coords1} into \re{masterbeta1}-\re{masterbeta2}. This, however, fails to yield a system of differential equations that can be solved exactly, due to the complicated dependence on the parameters $\beta$ and $\lambda$. It is nevertheless useful to integrate the equations numerically. In all numerical examples that follow, we impose the adiabatic conditions \re{adia1}-\re{adia2} and set the initial amplitude $\delta_c=10^{-5}$ at the initial time $N_0=-10$. The remaining initial conditions can be obtained from the relations \re{adia1}-\re{adia3}.



\begin{figure}
\centering
\includegraphics[width=7cm]{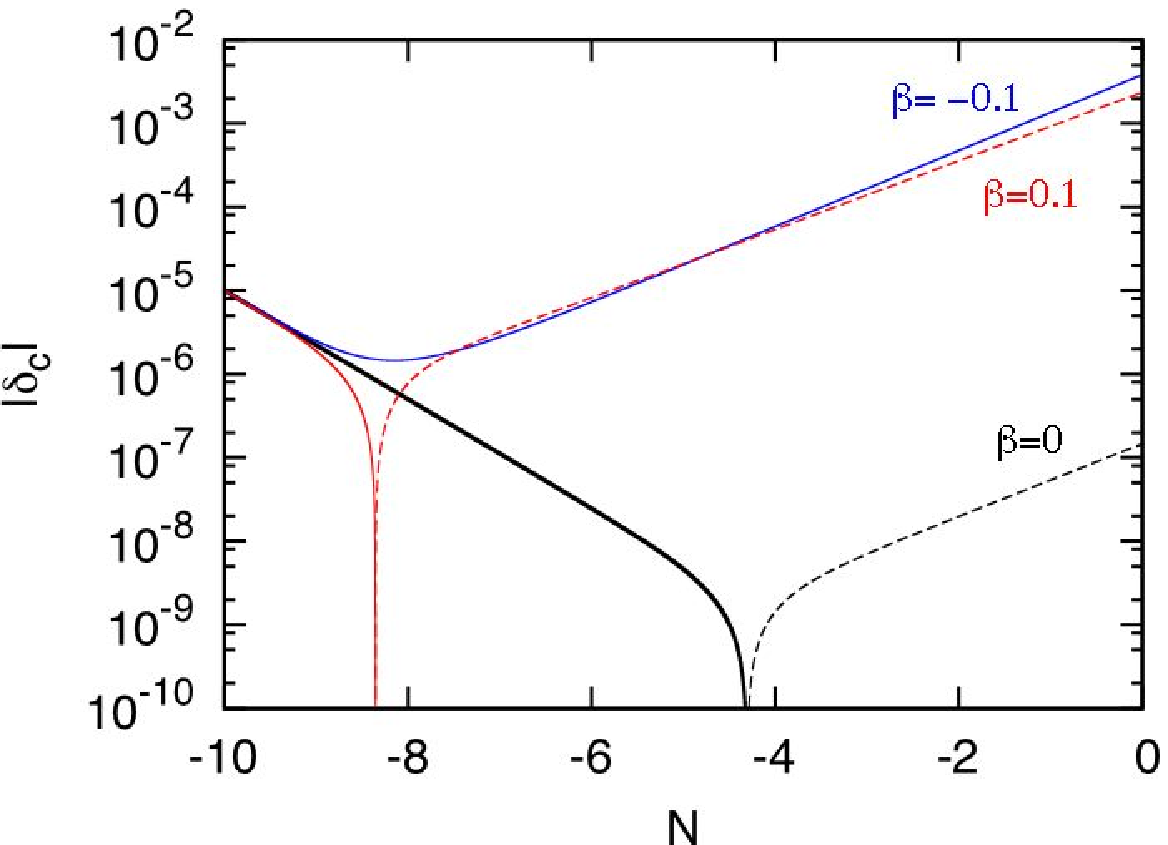}\ff\ff\includegraphics[width=7cm]{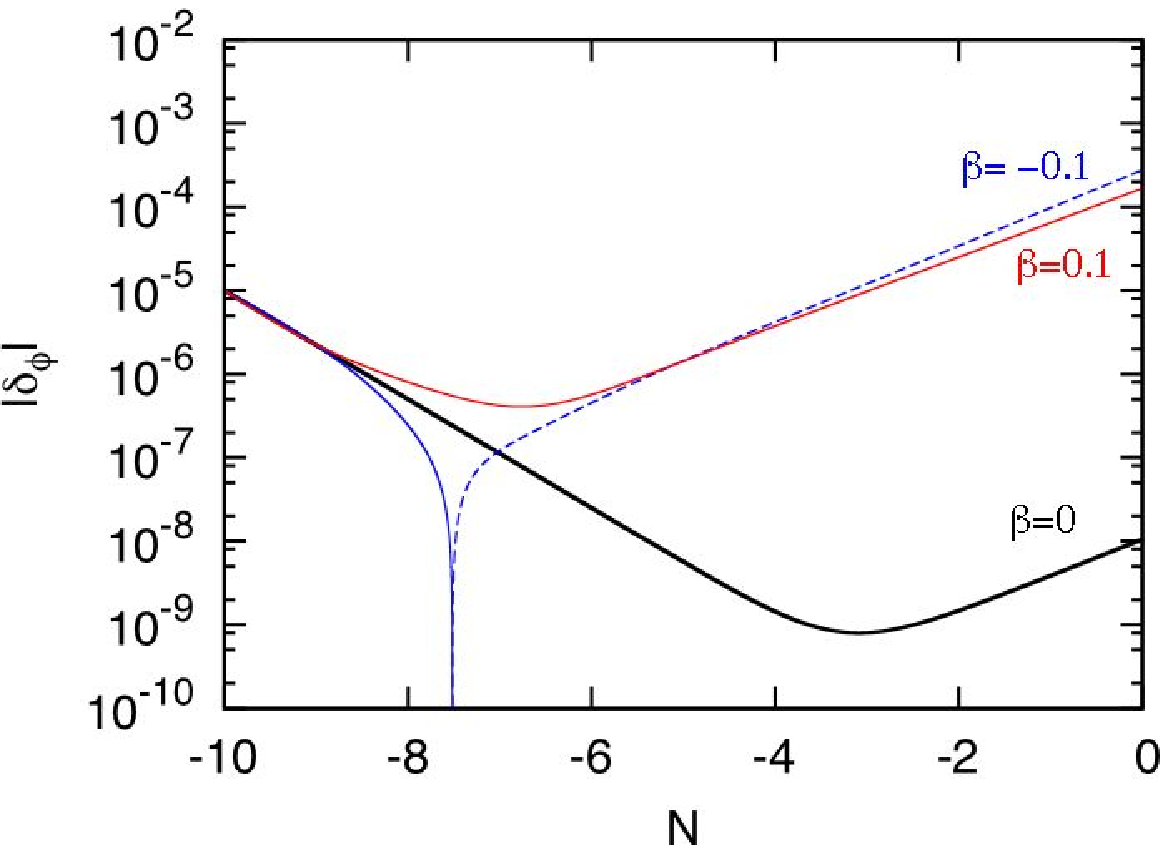}
\caption{Evolution of large-scale perturbations over $\Delta N=10$ during the scaling regime for the exponential potential with $\lambda=\sqrt5$, and interaction term of the form $Q=\beta H\rho_c$ (with $\beta = 0,\pm0.1$).  The panel on the left shows the evolution of the dark-matter density contrast $\delta_c$, while the panel on the right shows the quintessence density contrast $|\delta_\phi|$, showing similar trends. Dashed lines indicate negative values of the perturbations.}
\label{figbetascale}
\end{figure}

Figure \ref{figbetascale} shows the results of numerically integrating the dark-matter and quintessence density contrasts during the scaling regime of a large-scale mode (with $k/aH>10^3$ in the range $-10<N<0$). The adiabatic condition favours a large initial amplitude of the decaying mode. The gradient of the negative slope is approximately $-3/2$ in all cases (see Equation \re{scalinguncoupled}). The growing mode subsequently dominates. The gradient of the positive slope is approximately $1$ and increases slightly with  negative values of $\beta$.  Similarly, positive values of $\beta$ slows down the growth of the density contrasts. 

Dashed lines in the figure indicate negative values of the perturbations. Comparing the two panels in Figure \ref{figbetascale}, we see that at late times, clumping of dark energy is accompanied by dilution of dark matter (and vice versa). 
 

\begin{figure}
\centering
\includegraphics[width=7cm]{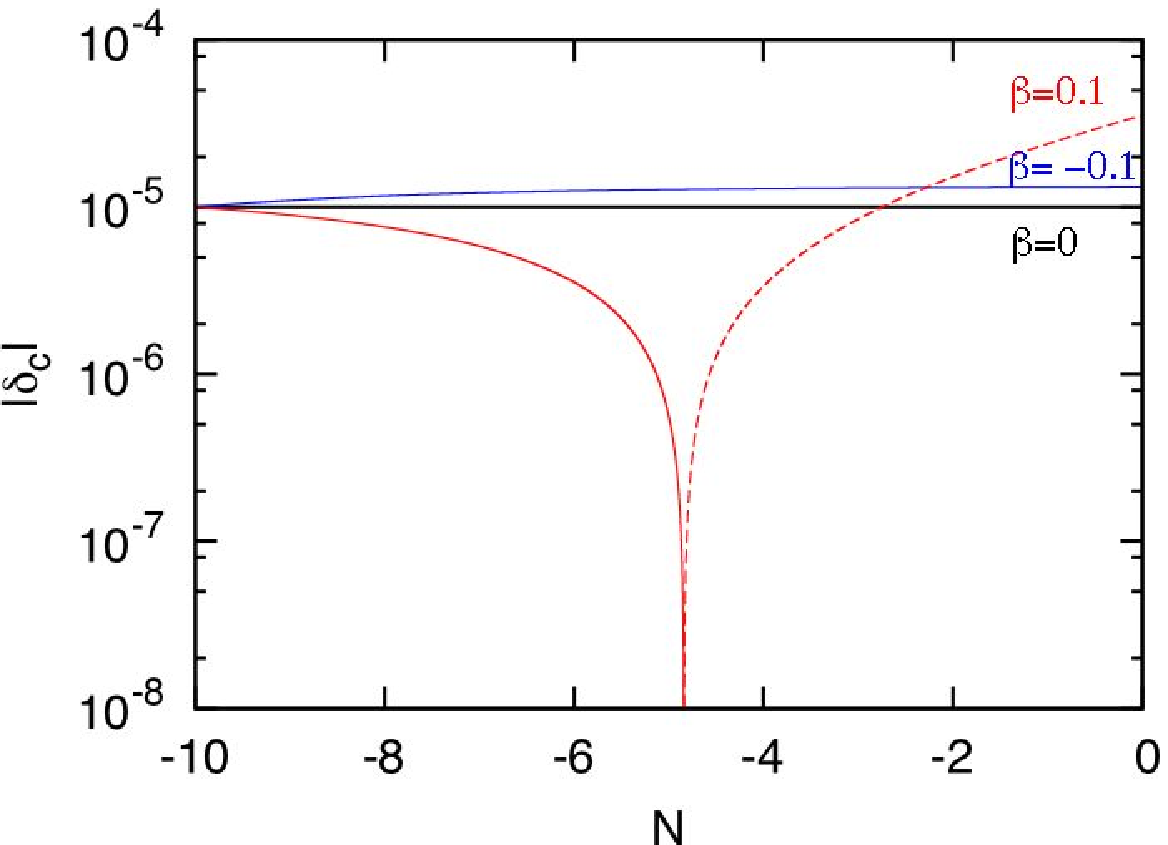}\ff\ff\includegraphics[width=7cm]{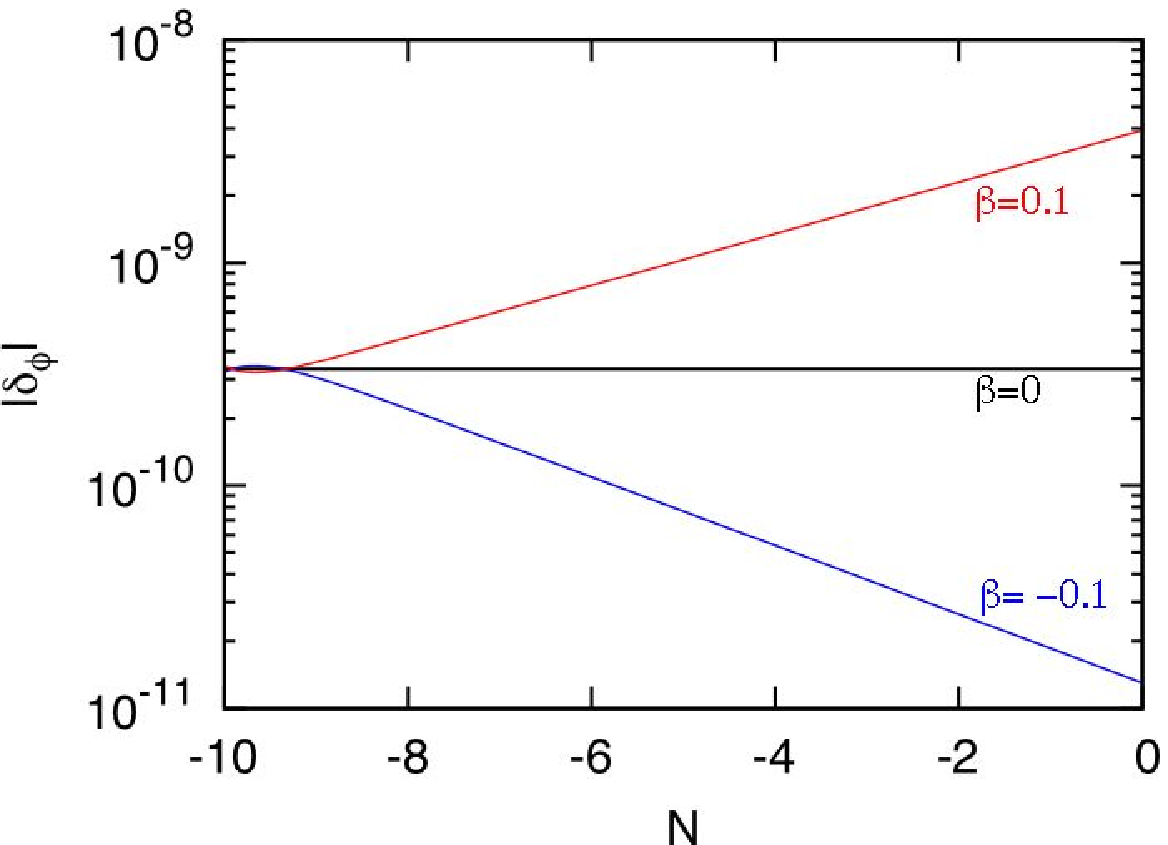}
\caption{Evolution of large-scale perturbations over $\Delta N=10$ during the quintessence-dominated regime for the exponential potential with $\lambda=0.01$, and interaction term of the form $Q=\beta H\rho_c$ (with $\beta = 0,\pm0.1$).  The panel on the left shows the evolution of the dark-matter density contrast $\delta_c$, while the panel on the right shows the quintessence density contrast $|\delta_\phi|$. Dashed lines indicate negative values of the perturbations. Asymptotic forms of these curves are given in the text. Note especially that $\delta_\phi$ is no longer constant if $\beta\neq0$.}
\label{figbetaflat}
\end{figure}
 

\subsubsection{Instability near $x=0$}

As we have seen, positive values of $\beta$ work in opposition to the motion of the quintessence. When the quintessence slows down sufficiently so that $x$ is close to zero, the background evolution must be handled with care, for $|dx/dN|\rightarrow\infty$ (this stems from the standalone $\gamma$ term in Equation \re{ode2}, with $\gamma=\beta u^2/2x$). This does not necessarily mean that the background is unstable, but it implies that any trajectory intersecting the $y$ axis stays on the $y$ axis. Putting $x=0$ into \re{ode3}, $y$ can be solved exactly as
\be y=(Ce^{-3N}+1)^{-1/2},\ee
with $C$ constant. This means that whenever a trajectory intersects the y axis, it then travels vertically towards $(0,1)$, which is the quintessence-dominated attractor. This behaviour can be partially seen in the left-most trajectory in Figure \ref{figbeta}. The numerical integration handles the trajectory so far as the intersection with the $y$ axis, at which point it breaks down. 

Whilst the background remains well-defined when the field comes to a stop, the perturbations are not so well-behaved. Almost every coefficient in \re{masterbeta1}-\re{masterbeta2} blows up as $x\rightarrow0$ implying that neither $\delta_c$ nor $\delta_\phi$ are stable at a linear level of perturbation. Hence, initial conditions in $(x,y,u)$ must be tuned so that $x$ never reaches 0. This sort of coupling is therefore not theoretically viable if quintessence is to alleviate the fine-tuning of initial conditions. This conclusion applies to any quintessence potential, and naturally extends to phenomenological interactions of the form \cite{olivares2}
\be Q\propto H(\rho_c+\rho_\phi).\ee
We note that a different type of instability in a similar model was identified in \cite{jussi} during the radiation-dominated era, in which the perturbations grow at extremely high rates.

\subsubsection{Flat limit}

For potentials with very gentle slope or flat `skater' models, the perturbations for large-scale modes can usually be solved analytically. For these models, the quintessence-dominated attractor exists and is stable whenever \cite{bohmer}
\be \lambda^2<3-\beta.\ee
By solving \re{masterbeta1}-\re{masterbeta2} and using the limit $\lambda\rightarrow0$ and $k\ll aH$, we find the usual scale-invariant mode $\delta_c=$ constant, and $\delta_c\sim e^{-2N}$ as before, plus two quintessence-induced modes. For $\beta<0$, we find the decaying modes
\be \delta_c\sim\delta\phi\sim e^{-{\beta+3\over2}N} (C_1\cos\theta N+C_2\sin\theta N),\lab{exact1}\ee 
where $C_i$'s are constant and $\theta=\sqrt{|5\beta^2+30\beta+9|}$. For $\beta>0$, we find 
\be \delta_c\sim\delta\phi\sim e^{-{\beta+3\over2}N} (C_3\cosh\theta N+C_4\sinh\theta N).\ee 
The salient feature of these solutions is that, for $\beta>0$, the growth rate of the mode
\be  \delta_c\sim\delta\phi\sim a^{-{\beta+3\over2}+\theta}\ee
surpasses that of the scale-invariant mode (since the exponent is positive for $\beta>0$). Hence, $\delta_c$ and $\delta_\phi$ could indeed grow during dark-energy-dominated era if $\beta>0$.

Figure \ref{figbetaflat} shows numerically evaluated perturbations for a few values of $\beta$, illustrating features outlined above. For these models, we use potentials with $\lambda=0.01$, as $\lambda=0$ would blow up the perturbation equations in the numerical program. The growing modes of the density contrasts when $\beta>0$ are clearly seen in both panels. The evolution of the quintessence density contrasts are also sensitive to $\beta$, as shown in the panel on the right. 




\subsection{Example III: $ Q=\sqrt{2\over3}b\kappa \dot\phi\rho_c$}

Dark-sector interaction of the form
\be Q=\sqrt{2\over3}b\kappa\dot\phi\rho_c,\ee
has been previously analysed in \cite{billyard,guo, gumjudpai, amendola,amendola2,wetterich}. This type of  interaction appears in some scalar-tensor theories with an interaction vector of the form
\be Q_\mu^{(c)}=\sqrt{2\over3}b\kappa T^{(c)}\phi_{;\mu},\lab{vecto}\ee
where $T^{(c)}=-\rho_c$ is the trace of $T^{(c)}_{\mu\nu}$. Here, we loosely refer to $\phi$ as quintessence although it is strictly more than just another form of cosmic energy density. In a scalar-tensor theory, $\phi$ is a mediator of gravitational interactions and can only be interpreted as part of the dark sector after a conformal transformation to the Einstein frame.

By perturbing the vector \re{vecto} and following the procedure in the previous section, we find the perturbed interaction to be
\be \delta Q^{(c)}=Q\bkt{\delta_c-2\alpha+{\delta\dot\phi\over\dot\phi}}.\ee
In terms of the dimensionless variables, the interaction and its perturbation are
\be\gamma &=& bu^2,\\
\delta\gamma &=& bu^2\bkt{\delta_c+{2\over\sqrt{6}}b(\kappa\delta\phi)}.\ee
Substituting the above into \re{master1}-\re{master2} gives the coupled differential equations governing the perturbations
\be {d^2\delta_c\over dN^2}&+&\bkt{{d\ln H\over dN}+2-2bx}{d\delta_c\over dN}+\bkt{2b^2-{3\over2}}u^2\delta_c = 2\sqrt{6}\bkts{{b\over3}+x-{b^2x\over3}}{d (\kappa\delta\phi)\over dN}+\bigg[2\sqrt6 b {d\ln H\over dN}-{16\over\sqrt6}b^3u^2\nn\\
&+&4\sqrt6 bx^2\bkt{1-{b^2\over3}}-{4\over\sqrt6}b^2x-{4b\over\sqrt6}\bkt{k\over aH}^2+2\sqrt6 b-\sqrt{6}b\eta y^2+3\lambda y^2(2b^2+1)\bigg](\kappa\delta\phi),\lab{masterb1}\\
{d^2(\kappa\delta\phi)\over dN^2}&+&\bkts{{d\ln H\over dN}+3+4bx}{d(\kappa\delta\phi)\over dN}+\bigg[3\eta y^2+\bkt{k\over aH}^2
+6b^2u^2+4b^2x^2-2\sqrt{6}b\lambda y^2\bigg](\kappa\delta\phi)\nn\\&=&\sqrt6 x{d\delta_c\over dN}-\sqrt{6}bu^2\delta_c. \lab{masterb2} \ee
Note that there is no longer any singularity near $x=0$.

\subsubsection{At the scaling solution}

\begin{figure}
\begin{center}
\includegraphics[width=5.2cm, height = 9.5cm, angle = -90]{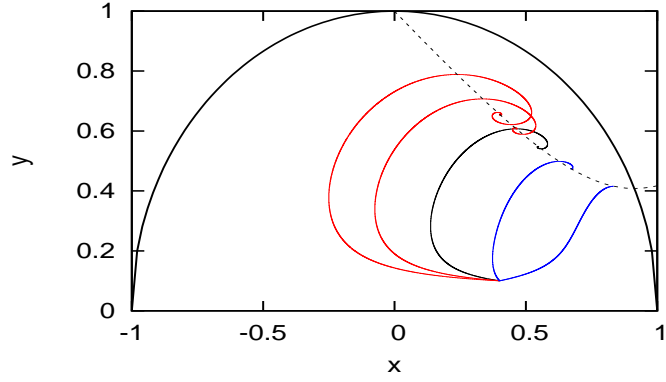}
\caption{Trajectories in the $x-y$ plane for the exponential potential with $\lambda=\sqrt5$ and interaction term $Q=\sqrt{2/3}b\kappa\dot\phi\rho_c$. From left to right, the trajectories within the semicircle correspond to $b=1, 0.5, 0, -0.5$ and $-1$. The dotted hyperbola is the locus of scaling solutions (Equation \re{locus}). In contrast with Figure \re{figbeta}, the region $x<0$ is now accessible to the trajectories since there is no singularity at $x=0$.}
\label{figb}
\end{center}
\end{figure}

For this model, the scaling solution exists whenever
\be -{3\sqrt{6}\over4}\leq Bb\leq{\sqrt6\over2}\bkt{B^2-3},\ee
where $B={\sqrt{6}\over3}b+\lambda$ \cite{bohmer}, with
\be x={3\over \sqrt6\lambda+2b}, \quad y ={\sqrt{4b^2+2\sqrt6\lambda b+9}\over \sqrt6\lambda+2b},\quad u= {\sqrt{6\lambda^2+2\sqrt6\lambda b -18}\over\sqrt6\lambda+2b},\quad {d\ln H\over dN}=-{3\sqrt{6}\lambda\over 2(\sqrt6\lambda+2b)}. \ee
Surprisingly, the locus of the scaling solutions remains the same hyperbola \re{locus}. The background dynamics in the phase space $(x,y)$ is shown in Figure \ref{figb}. Note that trajectories are now free to explore the entire semi-circle. If $b>0$, this type of interaction gives rise to an effective opposing force that slows down the field, just as in the previous case. In this case, the field can even roll `uphill' without encountering any instability. Similarly, we see that interaction with $b<0$ boosts the velocity of the field.


\begin{figure}
\begin{center}
\includegraphics[width=7cm]{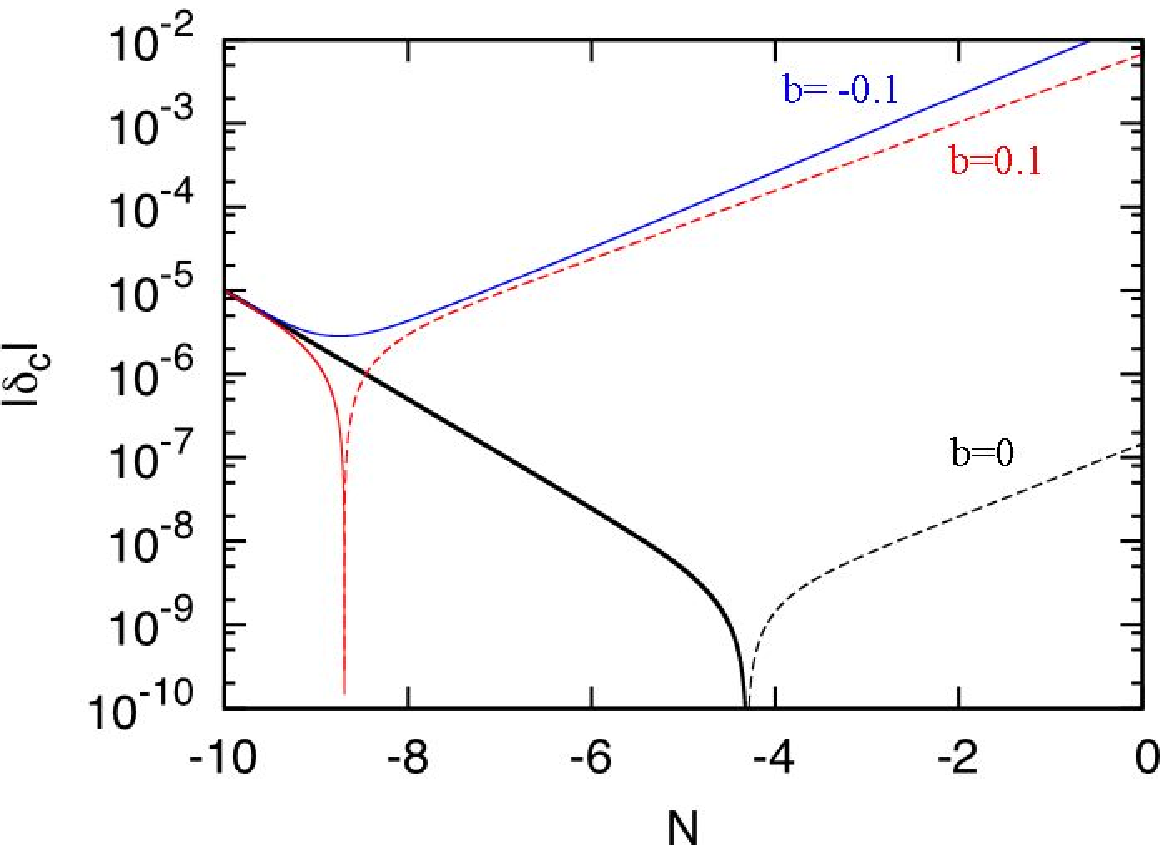}\ff\ff\includegraphics[width=7cm]{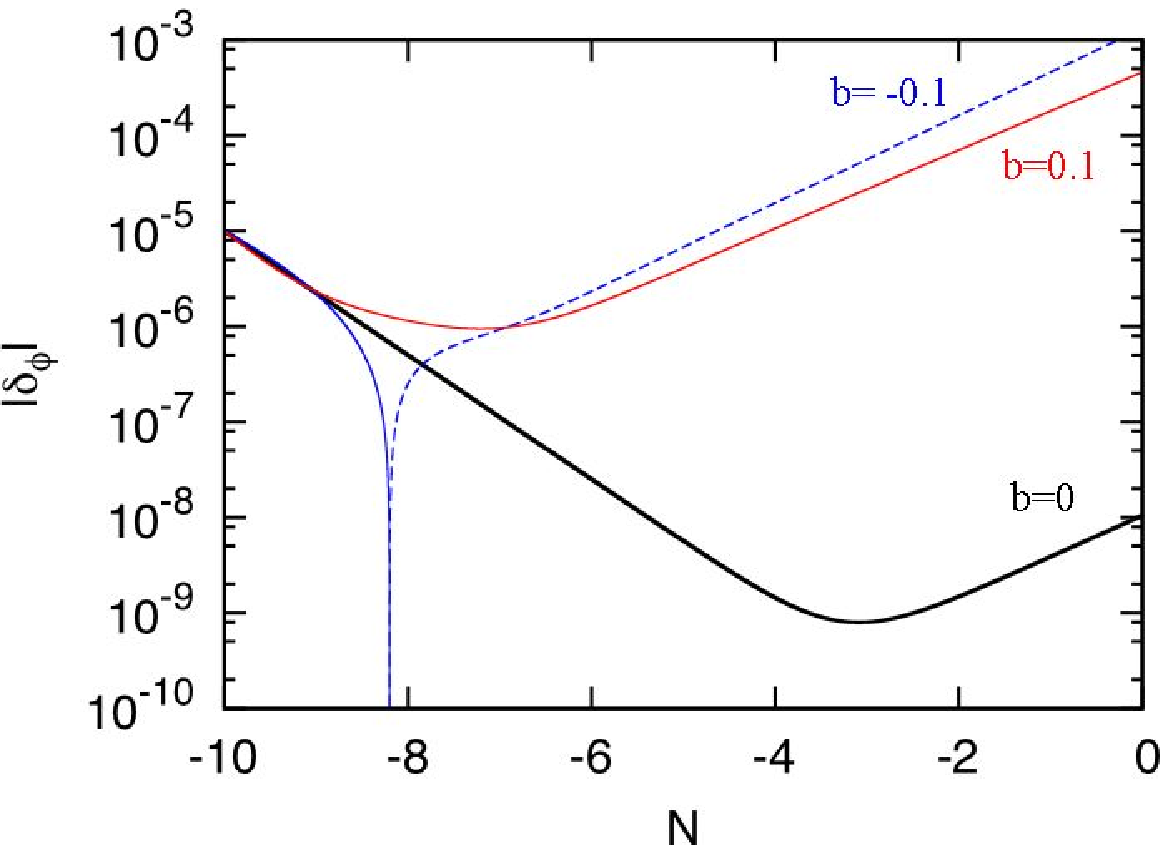}
\caption{Evolution of large-scale perturbations over $\Delta N=10$ during the scaling regime for the exponential potential with $\lambda=\sqrt5$ and interaction term $Q=\sqrt{2/3}b\kappa\dot\phi\rho_c$ (with $b=0,\pm0.1$). The panel on the left shows the evolution of density contrast, $\delta_c$, while the panel on the right shows the quintessence energy contrast, $\delta_\phi$. The dashed lines indicate negative values.}
\label{figbscale}
\end{center}
\end{figure}

\begin{figure}
\begin{center}
\includegraphics[width=7cm]{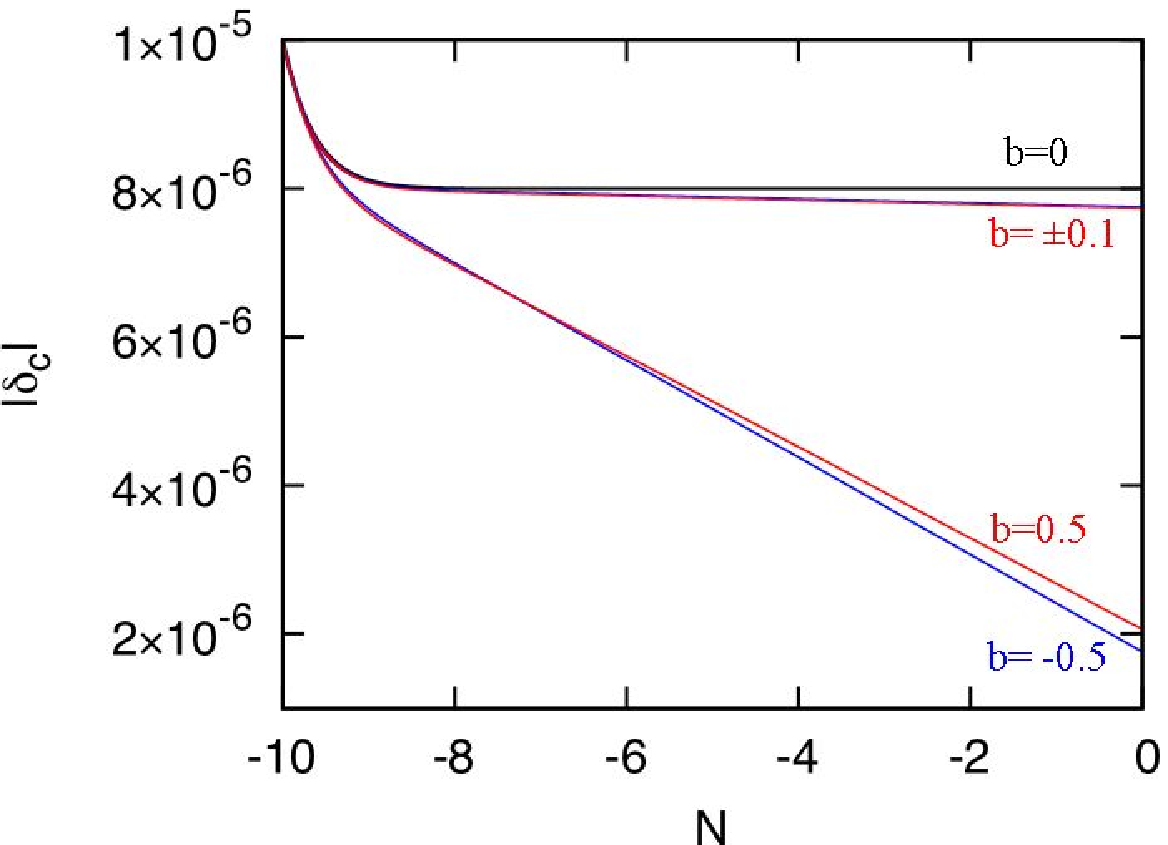}\ff\ff\includegraphics[width=7cm]{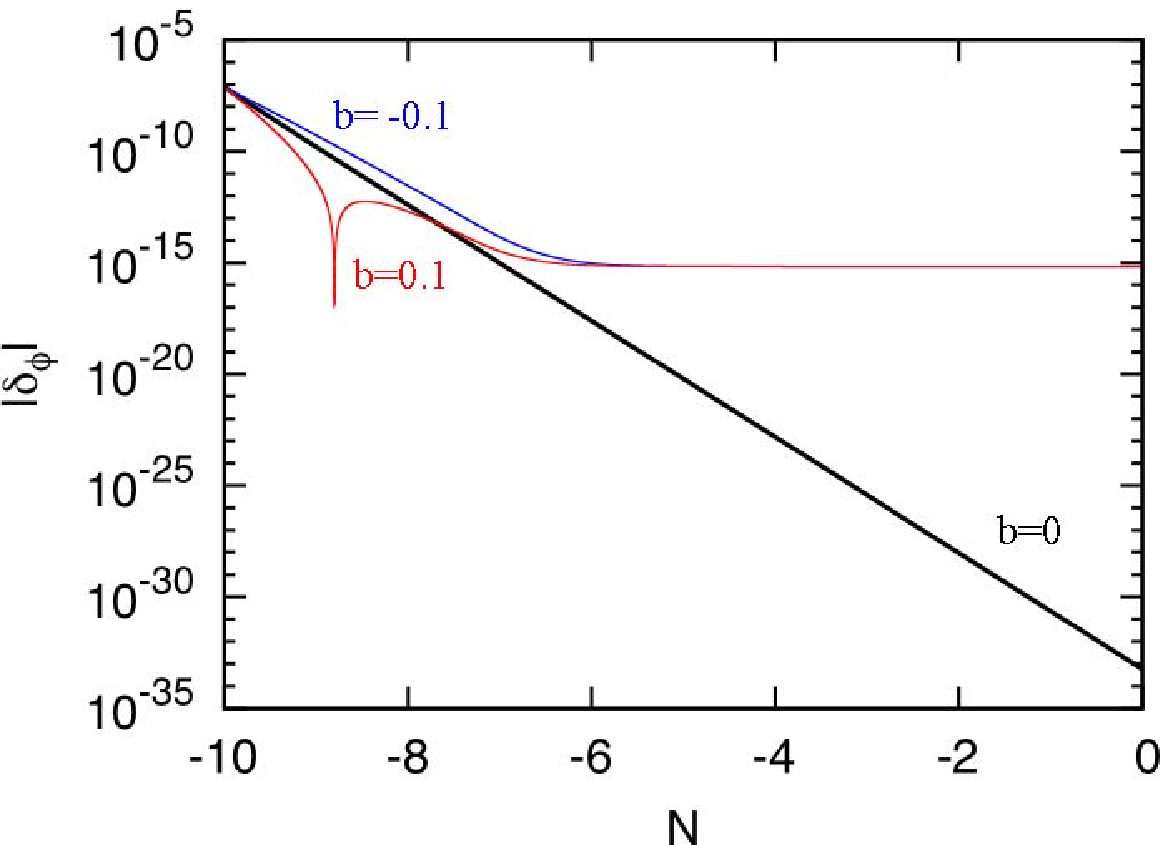}
\caption{Evolution of large-scale perturbations over $\Delta N=10$ during the quintessence-dominated regime for a flat potential $V=$constant ($\lambda=0$) and interaction term $Q=\sqrt{2/3}b\kappa\dot\phi\rho_c$. The panel on the left shows the evolution of density contrast $|\delta_c|$. Values of $b$ up to $\pm0.5$ are shown for comparison. The panel on the right shows the quintessence density contrast $|\delta_\phi|$. The asymptotic forms of these curves are discussed in the text. }
\label{figbflat}
\end{center}
\end{figure} 


Figure \ref{figbscale} shows the results of numerical integrations of the density contrasts $\delta_c$ and $\delta_\phi$. These results are similar to the previous case, \iee the growth rates of the perturbations increase with $b<0$ while the reverse occurs with $b>0$. The gradients of the lines representing decaying and growing mode are also approximately $-3/2$ and $1$ as in the previous case. However, the growth rates in this case respond more sensitively to changes in the interaction parameter.

\subsubsection{Flat limit}

The position and stability conditions for the quintessence-dominated attractor remain the same as in the previous coupling. Substituting in the coordinates \re{another} and using the large-scale and flat limits, the perturbation equations \re{masterb1}-\re{masterb2} can be solved exactly, with solutions
\be \delta_c = C_1+C_2N+C_3 e^{-2N},\quad \kappa\delta\phi={C_2\over \sqrt6b}+C_4 e^{-3N}, \lab{exact2}\ee 
where $C_i$ are constants and $b\neq0$. The case $b=0$ was covered in section \ref{nocouplingscaling}. Interestingly, there is a new linear mode in $\delta_c$, and, unlike the previous case, the quintessence-induced mode in $\delta\phi$ is simply constant. This implies that $\delta_c$ evolve linearly with $N$, while $\delta_\phi$ remains constant. This is in contrast with the case without interaction, in which $\delta_\phi$ decays exponentially to zero. 

Figure \ref{figbflat} illustrates the features outlined above. In the panel on the left, in which a linear scale is used, the adiabatic condition induces a linearly decaying $\delta_c$ for $b\neq0$, whilst $\delta_c$ remains constant for $b=0$. The difference between the two cases are only noticeable with larger values of the interaction, shown here up to $b=\pm0.5$. 

The panel on the right of Figure \ref{figbflat} shows the evolution of the large-scale quintessence perturbation $\delta_\phi$. The constant modes are clearly seen in the cases where $b\ne0$. Without any interaction, quintessence perturbations decay with slope $-3$, as expected from Equation \re{cubic}. Over $\Delta N=0$, the difference between the coupled and uncoupled cases grows to some 20 orders of magnitude. Consequently, the details of structure formation and large-scale CMB anisotropies in these cases will also be markedly different.

\begin{figure}
\begin{center}
\includegraphics[width=7cm]{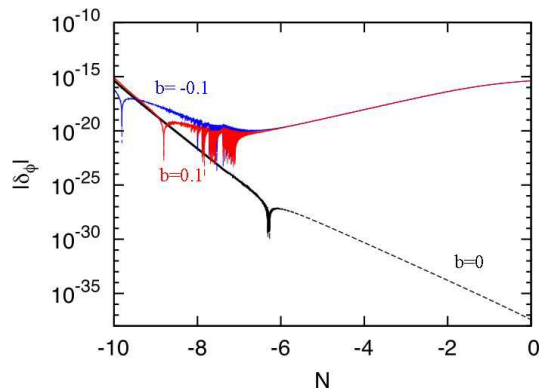}
\caption{Evolution of small-scale quintessence density contrast  during the quintessence-dominated regime for a flat potential $V=$constant ($\lambda=0$) and interaction term $Q=\sqrt{2/3}b\kappa\dot\phi\rho_c$. The dashed portions indicate negative values of $\delta_\phi$. }
\label{figbflatsmall}
\end{center}
\end{figure}

Moreover, such a vast difference in the magnitudes of dark-energy perturbations persists down to small scales, as shown in Figure \ref{figbflatsmall} (in which $k>aH$ in $-10\leq N\leq0$). Without interaction, $\delta_\phi$ decays to zero as before. For $b=\pm1$, $|\delta_\phi|$ is no longer constant on small scales, but oscillates briefly before a growing mode dominates\footnote{We find that $\delta_c$ is also enhanced on small scales, in agreement with the results of \cite{koivisto}. However, such large perturbations should be confirmed by a calculation which uses a nonlinear perturbation theory.}. This makes it possible for dark energy to clump on sufficiently small scales to have important astrophysical consequences, for instance, on the abundance of galaxy clusters and on gravitational lensing. 

\section{Conclusions}

In this paper, we have investigated the consequences of interactions between dark matter and dark energy in the form of quintessence. We have set up a general framework in which the density contrasts in dark matter and dark energy could be calculated given any quintessence potential $V(\phi)$ and dark-sector interaction $Q_\mu$. Our formalism is built on the phase-space method in which different cosmological histories correspond to trajectories in a finite space of dimensionless variables \re{dimless1}-\re{dimless2}. We have shown how the phase-space formalism may be extended so that perturbations in the densities of dark matter and dark energy may be calculated along the trajectories. This is particularly useful when combined with the knowledge of the attractor dynamics in the phase space. 

The key results of this paper are as follows:

\begin{itemize}
\item Using the results of Hwang and Noh \cite{hn}, we obtained a set of second-order coupled differential equations \re{master1}-\re{master2} describing the evolution of the perturbations in terms of dimensionless background variables [Equation \re{coeff}]. This is valid for any quintessence potential and interaction vector. We note that the perturbation $\delta\gamma$ in the interaction itself plays a crucial role as extra source terms in the coupled system.

\item We explained why adiabaticity is no longer conserved in the presence of interactions, even for large-scale perturbations [Equation \re{Sprime}].

\item Our formalism was applied to a number of scenarios involving the exponential potential. In particular, we studied the behaviour of perturbations as trajectories approach the scaling solution as well as the quintessence-dominated attractor in the case that the potential is almost flat. To test this technique, we set the interaction, $Q$, to zero and obtained analytic results that agree with those obtained by previous authors. 

\item We considered the interaction of the form  $Q=\beta H \rho_c$. At the background level, we found that $\beta>0$ effectively acts against the motion of the field, reducing its kinetic energy. Applying the perturbation equations, we found that the growth rates of perturbations change very slightly in the scaling regime. At the quintessence-dominated regime, we found analytic expressions for the growing and decaying modes for large-scale perturbations.

\item An instability associated with the interaction $Q=\beta H \rho_c$ was found. When the field velocity is close to zero, the perturbation equations blow up despite the fact that the background variables continue to be well-behaved. We believe that this is the first time that this type of instability has been identified. To avoid such an instability, the initial conditions must be tuned so that the field consistently rolls without stopping. To the author, such an extra tuning of initial conditions spoils the dynamical advantages of quintessence and makes this type of interaction unappealing.

\item We investigated the coupling $Q=\sqrt{2/3}b\kappa \dot\phi\rho_c$. We found that $b>0$ works against the motion of the field just as before. However, perturbative instabilities are absent in this model even when the field rolls `uphill'. In general, we conjecture that all models of dark-sector interaction with $Q=\mathcal{O}(\dot\phi^n)$ and  $n<1$, suffer from instability in the perturbations when the quintessence velocity is small. 


\item For $b\neq0$, dark-energy perturbations remain {constant} in the quintessence-dominated era instead of decaying away, even on small scales. Figure \ref{figbflatsmall} shows an example in which the difference in the dark-energy perturbations between the cases $b=0$ and $|b|=0.1$ grows to over 20 orders of magnitude. This opens the possibility that dark energy could clump on astrophysical scales. The implications for the CMB anisotropies, matter power spectra, galaxy cluster counts and gravitational lensing observations are interesting, and we shall investigate further into these issues.


\end{itemize}

\smallskip

As long as there is no conclusive evidence otherwise, one must remain open to the possibility that there may be interactions between the two most enigmatic components of the Universe. Nevertheless, as we have shown in this work, the mathematical details of dark-sector interactions can be challenging and must be handled with care. There are good prospects of identifying such interactions since a small interaction coefficient  may give rise to a dramatic increase in the magnitudes of dark-energy perturbations. It would be interesting and quite straightforward to apply our formalism to other types of potentials and interactions.

\bigskip 

\bigskip

\centerline{\bb{Acknowledgments}}

\bigskip

The author is grateful for many valuable comments from Timothy Clifton, Pedro Ferreira, Christopher Gordon and Tomi Koivisto. The author is supported by Lincoln College, Oxford.

\bibliographystyle{h-physrev3}
\bibliography{quintinter}

\end{document}